\documentclass[twocolumn,showpacs,unsortedaddress,superscriptaddress,showkeys,preprintnumbers,letterpaper]{revtex4-1}
\usepackage{soul}
\usepackage{acronym}
\usepackage{amsmath}
\usepackage{amssymb}
\usepackage{graphics}
\usepackage{subfigure}
\usepackage[usenames]{color}
\usepackage[normalem]{ulem}
\usepackage[bookmarksnumbered, bookmarksopen, breaklinks, colorlinks]{hyperref}
\pdfoutput=1

\newcommand{\CITA}{Canadian Institute for Theoretical Astrophysics, 60 St.
George Street, University of Toronto, Toronto, ON M5S 3H8, Canada}
\newcommand{\Perimeter}{Perimeter Institute for Theoretical Physics, Waterloo,
Ontario N2L 2Y5, Canada}
\newcommand{\AEI}{Albert-Einstein-Institut, Max-Planck-Institut f\"ur
Gravitationsphysik, D-30167 Hannover, Germany}
\newcommand{\Leibniz}{Leibniz Universit\"at Hannover, D-30167 Hannover, Germany}

\newcommand{\mat}[1]{{\bf #1}}

\newcommand*{\aye}{\mathrm{i}}

\newcommand{\Jmax}{{J_{\mathrm{max}}}}
\newcommand{\Kmax}{{K_{\mathrm{max}}}}
\newcommand{\Lmax}{{L_{\mathrm{max}}}}
\begin{document}


\title{Interpolating compact binary waveforms using the singular value decomposition}

\author{Kipp~Cannon}
\email{kipp.cannon@ligo.org}
\affiliation{\CITA}

\author{Chad~Hanna}  
\email{chad.hanna@ligo.org}
\affiliation{\Perimeter}

\author{Drew~Keppel}  
\email{drew.keppel@ligo.org}
\affiliation{\AEI}
\affiliation{\Leibniz}

\begin{abstract}
Compact binary systems with total masses between tens and hundreds of solar
masses will produce gravitational waves during their merger phase that are
detectable by second-generation ground-based gravitational-wave detectors.  In
order to model the gravitational waveform of the merger epoch of compact binary
coalescence, the full Einstein equations must be solved numerically for the
entire mass and spin parameter space.  However, this is computationally
expensive.  Several models have been proposed to interpolate the results of
numerical relativity simulations.  In this paper we propose a numerical
interpolation scheme that stems from the singular value decomposition.  This
algorithm shows promise in allowing one to construct arbitrary waveforms within
a certain parameter space given a sufficient density of numerical simulations
covering the same parameter space.  We also investigate how similar approaches
could be used to interpolate waveforms in the context of parameter estimation.
\end{abstract}

\maketitle
\acrodef{BNS}{binary neutron star}
\acrodef{CBC}{compact binary coalescence}
\acrodef{GW}{gravitational-wave}
\acrodef{PN}{post-Newtonian}
\acrodef{ROC}{Receiver Operator Characteristic}
\acrodef{SNR}{signal-to-noise ratio}
\acrodef{SPA}{stationary phase approximation}
\acrodef{SVD}{singular value decomposition}
\acrodef{IMR}{inspiral-merger-ringdown}
\acrodef{ASD}{amplitude spectral density}

\section{Introduction}

Searches for gravitational waves from binary black holes with total masses
between tens and hundreds of solar masses benefit from the complete model of
the gravitational waveform obtained by numerical
relativity~\cite{flanagan-1998-57, IMRreview2010}.  Numerically solving
Einstein's equations is now quite routine \cite{Pretorius:2005gq,
Campanelli:2005dd, Baker:2005vv, Pretorius:2007nq,Husa:2007zz, Hannam:2009rd,
Hinder:2010vn}, yet still computationally burdensome.  Reference
\cite{interplay2008} suggests that there is a finite density of numerical
simulations that would adequately cover the parameter space for certain
ground-based detectors.  In this work we explore this concept and extend the
numerical techniques presented in \cite{Cannon2010} and
\cite{Cannon2011manifold}, to interpolation of template waveforms using the
singular value decomposition.  This should allow for the construction of
gravitational waveforms with parameters between the numerically generated
waveforms.

The idea of interpolating gravitational waveforms has existed for over a
decade.  Interpolation of waveforms generated by post-Newtonian techniques was
described in \cite{Croce2000} and \cite{finn2005}.  In these references
analytic formulae for waveform interpolation were derived for particular PN
models.  Since 2005 the numerical relativity community has been generating a
substantial number of gravitational waveforms for the coalescence of binary
black holes \cite{Pretorius:2005gq, Campanelli:2005dd, Baker:2005vv,
Pretorius:2007nq,Husa:2007zz, Hannam:2009rd, Hinder:2010vn}.  Interpolation of
these waveforms has been accomplished primarily by (i) phenomenologically
fitting the simulations to closed-form expressions \cite{Ajith:2007qp,
Ajith:2007kx, Ajith:2009bn} or (ii) by numerically solving simpler
differential equations that capture the orbital dynamics combined with
numerical stitching of the ringdown phase \cite{Buonanno:2006ui, DN2007b,
Buonanno:2007pf, DN2007b, DN2008, Damour2009a, Buonanno:2009qa}.  In this work
we propose a different approach to interpolate a set of template waveforms.
This approach does not require careful tuning of fitting formulae or stitching
of waveforms and can be applied to any waveform set of sufficient density.

This paper is organized as follows.  First, we describe the technique for
interpolating waveforms via the singular value decomposition.  Second we apply
the technique to a set of waveforms containing all phases of the compact binary
coalescence, inspiral, merger and ringdown.  Third we discuss how these results
might be applied to the contruction of waveform families, ongoing gravitational
wave searches, and parameter estimation.

\section{Interpolation technique}

It was shown in \cite{Cannon2010} that the \ac{SVD} reduces the number of
template waveforms needed to search a given parameter space.  Additionally,
\cite{Cannon2011manifold} showed that arbitrary waveforms within the parameter
space could be reconstructed from the \ac{SVD} of a sufficiently dense template
bank.  Here we demonstrate a method to directly obtain approximate
reconstruction coefficients for arbitrary waveforms in the parameter space via
interpolation.  Consider a waveform family $\mat{h}(x, y)$ described by the
physical parameters $(x, y)$, and consider a set of these waveforms enumerated
by the index $\alpha$ drawn from a region of the parameter space,
$\mat{h}(x_{\alpha}, y_{\alpha})$.  Recall that a \ac{SVD} of these waveforms
allows each to be written as a linear combination of basis waveforms
$\mat{u}_\mu$ with coefficients $M_\mu(x_{\alpha}, y_{\alpha})$
\begin{equation} 
\mat{h}(x_{\alpha}, y_{\alpha}) = \sum_\mu M_\mu(x_{\alpha},
y_{\alpha})\,\mat{u}_\mu,
\end{equation}
where, in the formalism of \cite{Cannon2010} and \cite{Cannon2011manifold},
$M_\mu(x_\alpha, y_\alpha) := \sigma_\mu (v_{(2\alpha-1)\mu} + \aye
v_{(2\alpha)\mu})$ is the $\alpha$th combination of singular values
$\sigma_\mu$ and reconstruction coefficients $v_{(2\alpha-1)\mu}$ and
$v_{(2\alpha)\mu}$.  Recall also that waveforms with arbitrary physical
parameters from the same region of parameter space can also be reconstructed
using the basis vectors $\mat{u}_\mu$ by projecting the waveforms onto the
basis vectors to obtain the reconstruction coefficients---a computationally
expensive procedure,
\begin{equation}
\mat{h}(x, y) \approx \sum_\mu (\mat{h}(x,y) \cdot \mat{u}_\mu) \, \mat{u}_\mu.
\end{equation}
This can be used to define the arbitrary reconstruction coefficients as
\begin{equation}
M_\mu(x, y) = \mat{h}(x,y) \cdot \mat{u}_\mu.
\end{equation}
We seek the set of interpolated reconstruction coefficients $M_\mu'(x, y)$ that
can approximately reconstruct an arbitrary waveform from that region of
parameter space.

Compact binary gravitational waveforms with negligible effects from spin and
eccentricity are characterized by their component masses.  We will assume for
concreteness a two parameter family of waveforms $\mat{h}(x,y)$ where $x$ and
$y$ are $M$ and $q$, respectively, where $M=m_1+m_2$ is the total mass of the
system and $q=m_1/m_2$ is the mass ratio of the system.

Chebyshev polynomials of the first kind are known to be good for interpolation,
however other interpolation schemes are also possible.  We start with a set of
basis vectors $\mat{u}_j$ covering the desired region of parameter space. We
choose a net of points, scaled such that each dimension covers the interval
$[-1,1]$, located at the $\Jmax$th order Chebyshev nodes. For a single
dimension, these nodes occur at the locations
\begin{equation}
x_j = \cos \left( \pi \frac{j + \frac{1}{2}}{\Jmax+1} \right),
\end{equation}
where $j$ ranges from 0 to $\Jmax$. This choice of net reduces Runge's
phenomenon when used with the Chebyshev polynomials, which, for a single
dimension, are given as
\begin{equation}
T_J(x) = \frac{(x-\sqrt{x^2-1})^J + (x+\sqrt{x^2-1})^J}{2w},
\end{equation}
where $w=\sqrt{(1+\delta_{J0})(\Jmax+1)/2}$ is a normalization factor for the
polynomials and $\delta_{J0}$ is the Kroenecker delta. Both $x_j$ and $w$
depend on the choice of $\Jmax$, however for ease of notation we will leave
this implied.  The polynomials $T_J(x)$ satisfy the discrete orthogonality
condition
\begin{equation}
\sum_{j=0}^{\Jmax} T_I(x_j) T_J(x_j) = \delta_{IJ}.
\end{equation}
It is straightforward to extend this to higher dimensions.

In order to obtain the reconstruction coefficients for these locations, we
project waveforms from these locations onto the basis vectors.
From the values on this net, we interpolate to other positions in parameter
space using 2D-Chebyshev interpolation for each set of reconstruction
coefficients $M_\mu(x,y)$. Specifically, these values are projected onto the
Chebyshev polynomials
\begin{equation} \label{eq.cheby-coeff}
C_{KL\mu} = \sum_{k=0}^{\Kmax}\sum_{l=0}^{\Lmax}
T_K(x_k) T_L(y_l) M_{\mu}(x_k,y_l).
\end{equation}
This results in coefficients for the 2D-Chebyshev polynomials which can
be used to evaluate the interpolated reconstruction coefficients $M_\mu'(x,y)$
at other points in parameter space
\begin{equation} \label{eq.cheby-interp}
M_\mu'(x,y) = \sum_{K=0}^{\Kmax}\sum_{L=0}^{\Lmax} 
C_{KL\mu} T_K(x) T_L(y).
\end{equation}

In the next section we explore this approximation technique using gravitational
waveforms containing all three phases of binary coalescence, inspiral, merger
and ringdown.

\subsection{Reconstruction Errors}

Errors in reconstructing these waveforms come in two types:  errors due to
\ac{SVD} truncation, and errors due to reconstruction coefficient
interpolation. The truncation errors have previously been shown to take the
form
\begin{equation}
\left(\frac{\delta\rho(x,y)}{\rho(x,y)}\right)_{\rm trunc} =
\frac{1}{4}\sum_{\mu=N'+1}^{N} |M_{\mu}(x,y)|^2,
\end{equation}
where the sum is over the basis vectors that are discarded. The interpolation
errors have a similar form
\begin{equation}
\left(\frac{\delta\rho(x,y)}{\rho(x,y)}\right)_{\rm interp} =
\frac{1}{4}\sum_{\mu=1}^{N'} |M_{\mu}(x,y) - M_{\mu}'(x,y)|^2.
\end{equation}
It should be noted that here the sum is over the basis vectors that are kept
from the \ac{SVD}.  By setting the reconstruction coefficients with $\mu>N'$ to
zero, these can be combined into a single expression
\begin{equation}
\frac{\delta\rho(x,y)}{\rho(x,y)} = \frac{1}{4}\sum_{\mu=1}^{N}
|M_{\mu}(x,y) - M_{\mu}'(x,y)|^2.
\end{equation}
%

\section{Results}

We apply this procedure in two ways. In section~\ref{sec:whitenedwaveforms} we
investigate using this approach in the context of interpolating whitened
waveforms. This would be useful in the context of parameter estimation.
Specifically, one could obtain reconstruction coefficients that would be used
for constructing filter outputs associated with arbitrary points in parameter
space using the filter outputs from the \ac{SVD} basis vectors.

In section~\ref{sec:rawwaveforms}, we apply similar techniques to interpolate
raw waveforms. This is done in the context of waveforms one would receive from
numerical relativity simulations (i.e., time series of $\Psi_{2}(t) =
\partial_t^2 h_+(t) + \aye \partial_t^2 h_\times(t)$ that are restricted to lie
along lines of constant $M$). This approach could be taken to extend numerical
relativity waveform catalogs at greatly reduced computational cost.

\subsection{Whitened waveforms}
\label{sec:whitenedwaveforms}

We apply this procedure to non-spinning phenomenological \ac{IMR}
waveforms~\cite{Ajith:2009bn} with $M \in [60M_{\odot}, 80M_{\odot}]$, $q \in
[1, 10]$, whitened with an initial LIGO \ac{ASD}, and transformed to the time
domain. We generate a stochastic template bank~\cite{Harry2009} with $99\%$
minimal match for this range of parameters.  Since we are working with \ac{IMR}
waveforms, there is no well defined end of the waveform.  We choose to align
the whitened waveforms according to their peak amplitudes and compute
the \ac{SVD} basis vectors from these waveforms using the procedure described
in~\cite{Cannon2010}.  At this intermediate stage, if we were to look at how
the resulting reconstruction coefficients vary in parameter space, we would see
high frequency features that would be difficult to resolve with interpolation
without a high density interpolation net.

\begin{figure}
\includegraphics{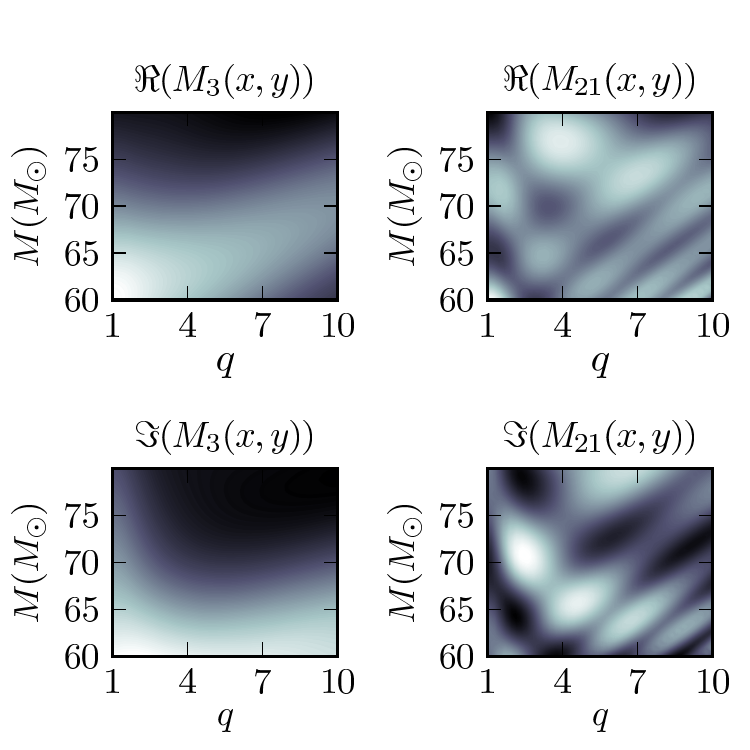}
\caption{Reconstruction coefficients as a function of $M$ and $q$ associated
with the $3^{\rm rd}$ and $21^{\rm st}$ basis vectors are shown in the left and
right columns, respectively. The top row shows the real part of the
reconstruction coefficients. The bottom row shows the imaginary part of the
reconstruction coefficients.}
\label{fig:recon}
\end{figure}

Fortunately, these features can be ameliorated by a complex rotation of the
input waveforms, which is equivalent to a complex rotation of the
reconstruction coefficients,
\begin{equation}
M_{\mu}(x,y) \rightarrow e^{-\aye \arg M_{1}(x,y)} M_{\mu}(x,y).
\end{equation}
This rotation is chosen such that $\Im\left[ M_{1}(x,y) \right] = 0$.
Fig.~\ref{fig:recon} shows the reconstruction coefficients associated with the
$3^{\rm rd}$ and $21^{\rm st}$ basis vectors after this complex rotation. The
smoothness of these reconstruction coefficients indicates that interpolation
should be possible.

\begin{figure}
\includegraphics{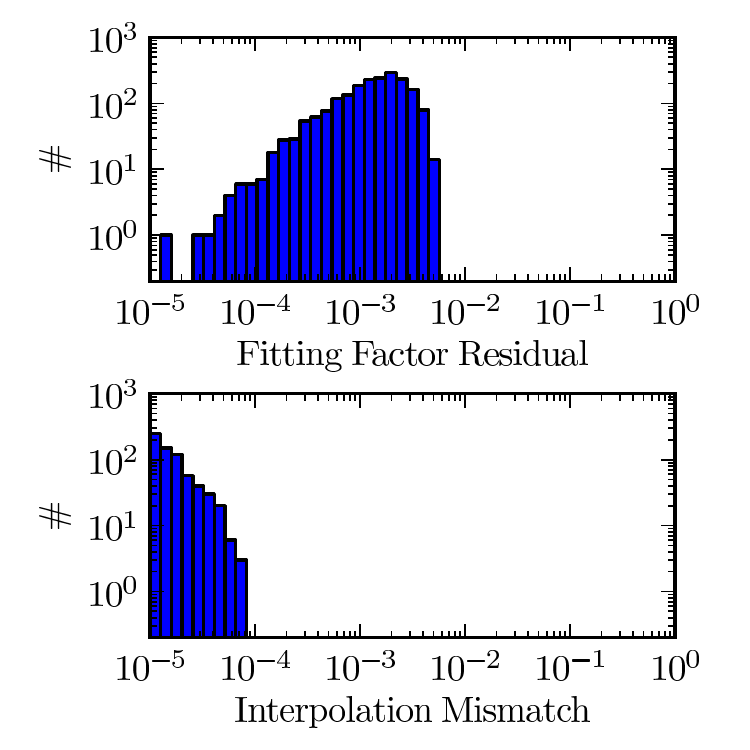}
\caption{The upper panel shows the fitting factor residual associated with
using the interpolation net as templates in a template bank. The lower panel
shows the interpolation mismatch for waveforms from with the parameter space.
The interpolation mismatch is more than an order of magnitude smaller than the
fitting factor residual.}
\label{fig:mismatch}
\end{figure}

In order to perform the interpolation, waveforms from the (12,12) order
2D-Chebyshev net are then projected onto these basis vectors to obtain the
interpolation coefficients, as described by \eqref{eq.cheby-coeff}, and rotated
as described above. $40 \times 40$ test waveforms from within the parameter
space, laid out in a grid, are used for computing mismatches between the
interpolated waveforms, given by \eqref{eq.cheby-interp}, and the original
waveforms. Fig.~\ref{fig:mismatch} compares the fitting factor residual, which
we define to be one minus the fitting factor, obtained from using the net
waveforms as templates with the interpolation mismatches associated with the
test waveforms. We see that the largest interpolation mismatch is more than an
order of magnitude smaller than the fitting factor residual from the net
waveforms.

\subsection{Raw waveforms}
\label{sec:rawwaveforms}

We apply similar techniques to waveforms of a type that would be provided by
numerical relativity simulations. Specifically, we use non-spinning
phenomenological \ac{IMR} waveforms~\cite{Ajith:2009bn} with $M \in
[60M_{\odot}, 80M_{\odot}]$, $q \in [1, 6]$, multiplied by $f^2$, which is
equivalent to taking two time-derivatives, and transformed to the time domain.
We use the same alignment and rotation techniques described in
section~\ref{sec:whitenedwaveforms} to prepare the waveforms for interpolation.

To generate the basis vectors that enclose this parameter space, we construct a
stochastic template bank with an additional constraint. The mass ratios of the
templates are restricted to take on values $q \in \{ q_j = 5x_j+1 | j \in [1,
6] \}$, where $x_j$ are the nodes associated with the 10th order Chebyshev
polynomial.

\begin{figure}
\includegraphics{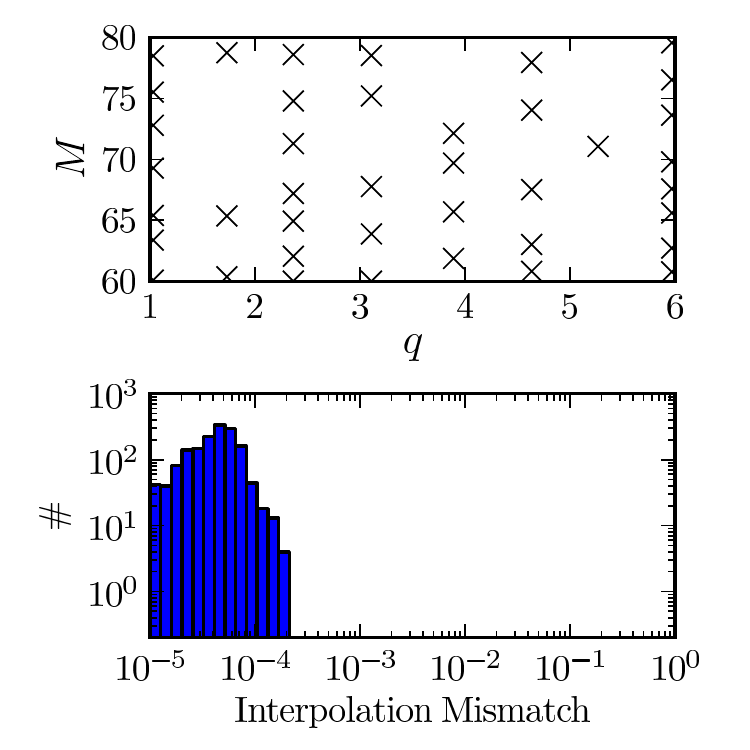}
\caption{The upper panel shows the locations for which waveforms were produced,
chosen by a stochastic template placement algorithm. These waveforms were used
in constructing the basis vectors enclosing this region of parameter space.
The lower panel shows the interpolation mismatch for waveforms from within the
parameter space. Waveforms interpolation accuracies are below a few times
$10^{-4}$.}
\label{fig:psi-mismatch}
\end{figure}

With the basis vectors in hand, we project the waveforms from an interpolation
net consisting of the (20,10) 2D-Chebyshev nodes onto the basis vectors to
obtain the reconstruction coefficients. These complex coefficients are rotated
as described above, and then used to obtain the interpolation coefficients.
Again, $40 \times 40$ test waveforms from within the parameter space, laid out
in a grid, are used for computing mismatches between the interpolated waveforms
and the original waveforms. We find comparable interpolation mismatches for
these non-whitened waveforms, shown in figure \ref{fig:psi-mismatch}, as for the
whitened waveforms.

\section{Conclusion}

Using the procedure described above, we have shown it is possible to produce
gravitational waveforms for arbitrary points in parameter space by
interpolating reconstruction coefficients from the \ac{SVD} of a set of
waveforms uniformly covering the space.

Results have been presented for both whitened waveforms, and raw waveforms.
The former could be useful in the context of parameter estimation associated
with \ac{CBC} \ac{GW} signals, which frequently uses Monte Carlo Markov Chain
methods to measure the likelihood ratio from many points in parameter space.
This requires the generation of the waveforms for each point in parameter space
and the overlap computation between the waveform and the data. Using the
interpolated reconstruction coefficients, the same computation could be
approximately performed with generating a subset of the total waveforms,
reconstructing the overlap by appropriately recombining the filter outputs from
the basis vectors. The latter could be used to accurately interpolate waveforms
that are computationally costly to produce, as is the case for numerical
relativity waveforms.

For future work, these techniques should be expanded to include additional
dimensions of parameter space (e.g., binary object spin parameters). In
addition, other interpolation schemes that use equispaced or random points in
parameter space might be found to be favorable for different applications.  We
also note that these techniques could be applied to other gravitational
waveforms such as supernova waveforms where singular value decomposition has
also been applied~\cite{heng2008}, or where other methods have been used to
reduce the rank of the parameter space~\cite{field2011}.

\acknowledgments

We would like to thank Christian R\"{o}ver and Ilya Mandel for discussions and
comments related to this work. Research at Perimeter Institute is supported
through Industry Canada and by the Province of Ontario through the Ministry of
Research \& Innovation. KC was supported by the National Science and
Engineering Research Council, Canada.  DK was supported from the Max Planck
Gesellschaft.  This work has LIGO document number {LIGO-P1100101-v2}.

\bibliography{references}
\end{document}